\documentclass[aip,jcp,reprint,amsmath,amssymb]{revtex4-2}
\usepackage{graphicx}
\usepackage[suffix=]{epstopdf}
\usepackage{natmove}
\newcommand{\vv}[1]{\mathbf{#1}}
\renewcommand{\d}[1]{\ensuremath{\operatorname{d}\!{#1}}}

\begin{document}
\title{Inverse design of drying-induced assembly of multicomponent colloidal-particle films using surrogate models}

\author{Mayukh Kundu}
\affiliation{Department of Chemical Engineering, Auburn University, Auburn, AL 36849, USA}

\author{Michaela Bush}
\affiliation{Department of Chemical Engineering, Auburn University, Auburn, AL 36849, USA}

\author{Chris A. Kieslich}
\affiliation{Wallace H. Coulter Department of Biomedical Engineering, Georgia Institute of Technology, Atlanta, Georgia 30332, United States}

\author{Michael P. Howard}
\email{mphoward@auburn.edu}
\affiliation{Department of Chemical Engineering, Auburn University, Auburn, AL 36849, USA}

\begin{abstract}
The properties of films assembled by drying colloidal-particle suspensions depend sensitively on both the particles and the processing conditions, making them challenging to engineer. In this work, we develop and test an inverse-design strategy based on surrogate modeling to identify conditions that yield a target film structure. We consider a two-component hard-sphere colloidal suspension whose designable parameters are the particle sizes, the initial composition of particles, and the drying rate. Film drying is simulated approximately using Brownian dynamics. Surrogate models based on Gaussian process regression (GPR) and Chebyshev polynomial interpolation are trained on a loss function, computed from the simulated film structures, that guides the design process. We find the surrogate models to be effective for both approximation and optimization using only a small number of samples of the loss function. The GPR models are typically slightly more accurate than polynomial interpolants trained using comparable amounts of data, but the polynomial interpolants are more computationally convenient. This work has important implications not only for designing colloidal materials but also more broadly as a strategy for engineering nonequilibrium assembly processes.
\end{abstract}

\maketitle

\section{Introduction}
Films made of colloidal particles have numerous technological applications, including latex paints \cite{keddie:1997,keddie:2010, routh:2013}, abrasion- \cite{tinkler:2021, kargarfard:2021} or bacteria-resistant \cite{dong:acs:2020} coatings, and lithium-ion battery electrodes \cite{zhang:2022}. One strategy for preparing these films is to self-assemble a suspension of colloidal particles into a solid by evaporating the solvent. Different particle structures in the film can be achieved depending on the processing conditions, and these structures in turn dictate the film's properties \cite{routh:2013}. For example, a binary mixture of small and big colloidal particles can spontaneously stratify into a layered structure with the small particles accumulated near the film's air--solvent interface and the big particles accumulated near the substrate \cite{nikiforow:2010,trueman:lng:2012,atmuri:2012, fortini:2016, makepeace:2017, liu:2018}; the stratified particle distribution can be used to modulate properties such as anti-microbial activity and abrasion resistance\cite{dong:acs:2020, tinkler:2021}. However, stratification may not occur if the drying rate is too slow or too fast \cite{tatsumi:2018, liu:2019, schulz:2022, fernando:2025}, and other structures may also form depending on the particle sizes and interactions\cite{MartinFabiani:2016fj, tang:2019, tinkler:2022, liu:2023}. Processing conditions can hence be used to engineer the final structure of a colloidal-particle film, but the relationship between the two remains incompletely understood.

Efforts to study self-assembly of colloidal particles in drying films have largely focused on ``forward'' strategies using physical \cite{trueman:lng:2012,fortini:2016, makepeace:2017, carr:2018, schulz:2021} or computational \cite{routh:2004, trueman:jcis:2012,howard:lng:2017,fortini:2017, tang:2019, chun:2019, yoo:2020} experiments, as well as approximate theories \cite{sear-warren:2017,sear:2018}, to create state diagrams mapping the film's structure onto key process parameters such as particle size, initial particle volume fraction, and drying rate (P\'{e}clet number) \cite{cardinal:2010, zhou:prl:2017, schulz:2018, schulz:2021, fernando-gunawardana-thilanka:2025}. However, these diagrams can be time- and resource-intensive to construct due to the number of combinations of parameters that must be considered. They also lose information about the specific structures that are formed (e.g., not all films that stratify have the same extent of particle segregation), which limits their use for engineering films with specific structures. In this context, ``inverse'' strategies that can efficiently identify parameters for assembling a film with a designed structure are attractive \cite{ferguson:2017, jackson:2019, sherman:2020}.

Inverse design strategies for colloidal self-assembly typically use a loss function to measure how closely the structure and/or properties that are produced by a given set of particle and processing parameters match desired values \cite{sherman:2020}. These parameters are then adjusted to minimize the loss function. Direct optimization of the loss function can be costly because each evaluation usually requires an experiment or simulation. Surrogate modeling is a technique used in a variety of engineering fields to help overcome this difficulty \cite{boukouvala:2013, amaran:2016, boukouvala:2016, bhosekar:2018}. Surrogate models approximate complex functions that may not have a known form \cite{jones:1998, azarhoosh:2025} by sampling it for some combinations of inputs, then training a simpler function on these samples to approximate the complex function on its entire input space, including points that were not sampled \cite{jones:1998, caballero:2008, regis:2014, azarhoosh:2025}.

Typically, there are two critical decisions for constructing a surrogate model: where to sample and what functional form to use. Here, we focus on two approaches that are known to be effective with small amounts of training data \cite{kieslich:2018, petix:2023, fakhraei:2025}: Gaussian process regression (GPR) and Chebyshev polynomial interpolation. GPR is a probabilistic strategy to approximate a function, including an estimate of prediction uncertainty \cite{williams:1995, rasmussen:2004, williams:2006}. GPR does not require a specific sampling strategy but stochastic techniques  \cite{mckay:2000, nishant:2018} such as Latin hypercube sampling \cite{stein_large_1987} are common, and it is also naturally compatible with adaptive sampling strategies \cite{wang:2018}. Chebyshev polynomial interpolation \cite{Press:2002} uses multivariate basis functions and sample points that are constructed by taking dense or sparse products of sets of univariate polynomials and corresponding sample points \cite{judd:2014, kieslich:2018, petix:2023, fakhraei:2025}. There are practical computational differences in how each type of surrogate model is trained and evaluated \cite{smolyak:1963, judd:2014, kieslich:2018, williams:2006}, and they may perform differently due to the differences in their sampling and functional forms. Chebyshev polynomial interpolation is known to perform favorably compared to GPR for surrogate-based optimization \cite{kieslich:2018}, but to our knowledge, neither has been previously applied to the problem we are considering.

In this work, we investigate the suitability of surrogate models for designing the drying-induced assembly of multicomponent colloidal-particle films. We model a bidisperse hard-sphere suspension using Brownian dynamics (BD) simulations, and we seek to design the film composition with respect to the particle sizes, the drying rate, and the initial particle volume fractions. We compare the accuracy of three different surrogate-modeling strategies (GPR, dense and sparse Chebyshev polynomial interpolation) for representing the loss function as well as for identifying suitable process parameters to obtain a particular structure. We find that all three strategies can be used effectively, with sparse Chebyshev polynomial interpolation striking a good balance between accuracy and computational considerations.

The rest of the article is organized as follows. We describe the simulation model for drying in Sec.~\ref{sec:model}, the design problem in Sec.~\ref{sec:objective}, and the surrogate modeling strategies in Sec.~\ref{sec:surrogate-models}. We then critically assess the suitability of the surrogate models for approximating the  loss function used for design in Sec.~\ref{sec:results}. We finally summarize our findings and provide an outlook on the use of surrogate models to design drying-induced assembly processes in Sec.~\ref{sec:conclusions}.

\section{Simulation model}
\label{sec:model}
We simulated a two-component drying colloidal suspension at constant temperature $T$. The suspension consisted of small (S) and big (B) nearly hard particles with diameters $d_{\rm S}$ and $d_{\rm B}$, respectively. The interactions between pairs of particles were modeled using the repulsive Weeks--Chandler--Andersen potential \cite{Weeks:1971},
\begin{equation}
\beta u(r_{ij}) = \begin{cases}
\displaystyle 4 \left[\left(\frac{d_{ij}}{r_{ij}}\right)^{12}-\left(\frac{d_{ij}}{r_{ij}}\right)^6\right] + 1,& r_{ij} \le 2^{1/6} d_{ij} \\
0,& {\rm otherwise}
\end{cases},
\label{eq:wca}
\end{equation}
where $r_{ij}$ is the distance between the centers of particles $i$ and $j$, $d_{ij}$ is the arithmetic mean of the particle diameters, and $\beta = 1/(k_{\rm B}T)$ with $k_{\rm B}$ being the Boltzmann constant. The particles were suspended above a purely repulsive substrate (wall) with normal in the $z$ direction. Particle interactions with the substrate were modeled using a repulsive Lennard-Jones 9--3 potential
\begin{equation}
\beta u_{\rm W}(z_i) = \begin{cases}
\displaystyle \varepsilon_i \Bigg[ \frac{2}{15}\left(\dfrac{d_{{\rm W},i}}{z_i}\right)^9 - \left(\frac{d_{{\rm W},i}}{z_i}\right)^3  \\
\displaystyle \quad \quad+ \frac{\sqrt{10}}{3} \Bigg] , \quad z_i \le \left(\dfrac{2}{5}\right)^{1/6} d_{{\rm W},i} \\
0, \quad {\rm otherwise}
\end{cases},
\label{eq:ljwall}
\end{equation}
where $z_i$ is the $z$ position of particle $i$, $d_{{\rm W},i}$ is the contact diameter of the particle with the substrate, and $\varepsilon_i$ sets the strength of the repulsion. The film's liquid--air interface was modeled using a repulsive harmonic potential\cite{howard:lng:2017, fortini:2016, pieranski:prl:1980},
\begin{equation}
\beta u_{\rm LV}(z_i,t) = \begin{cases}
\displaystyle \frac{\kappa_i}{2}\left[z_i-H(t)+\frac{d_i}{2}\right]^2,& z_i > H(t)-\dfrac{d_i}{2} \\
0,& {\rm otherwise}
\end{cases},
\label{eq:slv}
\end{equation}
where $\kappa_i$ sets the strength of the repulsion for particle $i$, $d_i$ is the particle's diameter, and $H(t)$ is the position of the interface at time $t$. We assumed the interface receded with a constant velocity $v$ from an initial position $H_0$, so $H(t) = H_0 - vt$. The shift $-d_i/2$ ensured the particles remained fully immersed below this interface \cite{tang:2018}.

Particle motion was simulated using BD with free-draining hydrodynamic interactions \cite{Allen:2017,ermak:jcp:1978}. The small and big particles had diffusion coefficients $D_{\rm S}$ and $D_{\rm B}$, respectively, with $D_{\rm B}/D_{\rm S} = d_{\rm S}/d_{\rm B}$ to be consistent with the Stokes--Einstein relation. We note that this model may not quantitatively predict the film's structure due to its neglect of solvent backflow and pairwise hydrodynamic interactions between particles \cite{sear-warren:2017, statt:2018, howard:2020, kundu:2025}; however, it should still produce structures qualitatively similar to those observed in experiments, including small-on-top stratified layers \cite{fortini:2016, howard:lng:2017, liu:2019, yetkin:2023}, at less computational cost than models that do incorporate these hydrodynamic effects \cite{tang:lng:2018, tang:2019, chun:2019, park:2022, kundu:2025}. As such, our free-draining BD simulations are suitable for exploring the use of surrogate models to design self-assembly in drying films, but they likely do not make quantitative predictions that translate directly to experiments. 

We used LAMMPS (2 August 2023 update 3) to perform all simulations\cite{thompson_lammps:2022}. We defined $k_{\rm B} T$ as the unit of energy, the diameter of the small particles $d_{\rm S}$ as the unit of length, and the characteristic time for a small particle to diffuse its own diameter, $\tau_{\rm S} = d_{\rm S}^2/D_{\rm S}$, as the unit of time. Hence, the diameter and diffusion coefficient of the small particles were effectively fixed, but the diameter and diffusion coefficient of the big particles were varied. For the substrate interactions, we used $d_{{\rm W},i} = (d_{\rm S} + d_i)/2$ for type $i$, $\varepsilon_{\rm S} = 1$ for the small particles, and $\varepsilon_{\rm B} = 100$ for the big particles; the larger value of $\varepsilon_{\rm B}$ was required to reduce the penetration of big particles into the substrate in some simulations. For the solvent interface, we used $\kappa_{\rm S} = \kappa_{\rm B} = 100\,d_{\rm S}^{-2}$. The simulation timestep was $10^{-5}\,\tau_{\rm S}$. 

We used an orthorhombic simulation box that was periodic in the $x$ and $y$ directions, with each of these directions having length $L = 100\,d_{\rm S}$. The initial particle configuration was prepared by randomly placing particles in sites of face-centered cubic lattices with lattice constants of approximately $\sqrt{2} d_i$ to achieve the desired initial volume fractions $\phi_{0,{\rm S}}$ and $\phi_{0,{\rm B}}$, where $\phi_{0,i} = N_i \pi d_i^3/(6L^2H_0)$ and $N_i$ is the number of particles for type $i$.  A padding of $d_i/2$ was included between the lattices and the film's liquid--air interface and substrate so that no particles overlapped with either. Big particles were first placed in their lattice, then small particles were added to sites in their lattice that did not overlap with the big particles.  A short equilibration of $1\,\tau_{\rm S}$ was run, then the film was dried at the specified interface velocity $v$ until the desired final film height $H_1$ was achieved. This simulation protocol mimics casting of a well-mixed suspension followed immediately by drying \cite{routh:2013}.

\begin{figure}
    \centering
    \includegraphics{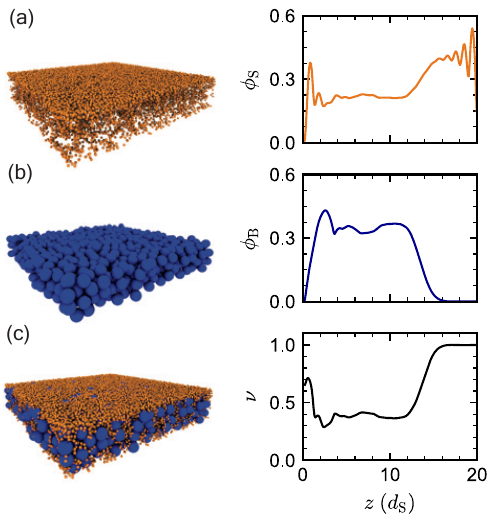} 
    \caption{Example simulation of film consisting of small (orange) and big (blue) particles. The small particles are stratified on top of the big particles, which is highlighted by showing (a) small particles only, (b) big particles only, and (c) both particles together. The corresponding position-dependent volume fractions (a) $\phi_{\rm S}(z)$ and (b) $\phi_{\rm B}(z)$ as well as (c) composition by volume $\nu$ are shown alongside each representation.}
    \label{fig:loss-schematic}
\end{figure}

We characterized the dried film structure using its composition by volume in the final particle configuration recorded in the simulations. First, we computed the position-dependent number density $\rho_i(z)$ of each type $i$ for $0 \le z \le H_1$ using histograms with bin spacing $0.1\,d_{\rm S}$. These number densities were converted to position-dependent volume fractions $\phi_i(z)$ by numerically convolving $\rho_i(z)$ with the volume of a sphere with diameter $d_i$ [Fig.~\ref{fig:loss-schematic}(a)--(b)] \cite{kundu:2022}. Last, we computed the position-dependent composition by volume, $\nu(z) = \phi_{\rm S}(z) / [\phi_{\rm S}(z) + \phi_{\rm B}(z)]$ [Fig.~\ref{fig:loss-schematic}(c)]. With this definition, $\nu$ is one when the volume is locally filled only with small particles and zero when it is filled only with big particles.

\section{Design problem}
\label{sec:objective}
For our bidisperse suspension, the particle and processing parameters that can nominally be used to manipulate the self-assembled film structure are: the particle diameters $d_{\rm S}$ and $d_{\rm B}$, the solvent (which determines $D_{\rm S}$ and $D_{\rm B}$), the initial film height $H_0$, the initial volume fractions $\phi_{0,{\rm S}}$ and $\phi_{0,{\rm B}}$, and the drying rate (interface speed $v$). Not all of these parameters are independent, however, and it is helpful to reduce the number of parameters considered in the design process using physical knowledge or engineering constraints. For this purpose, we assumed the initial film height $H_0$ and total initial volume fraction $\phi_0 = \phi_{0,{\rm S}} + \phi_{0,{\rm B}}$ were fixed at $H_0 = 50\,d_{\rm S}$ and $\phi_0 = 0.20$. These constraints mimic use of a concentrated dispersion and a thin initial coating \cite{butt:2022}. We also know that the P\'{e}clet numbers ${\rm Pe}_{\rm S}$ and ${\rm Pe}_{\rm B}$ for the small and big particles, defined as ${\rm Pe}_i = v H_0/D_i$ for type $i$, are important dimensionless numbers for describing the drying process \cite{schulz:2018, fernando-gunawardana-thilanka:2025}. Significantly different structures are usually observed when ${\rm Pe}_{\rm S}$ varies by orders of magnitude, and nonequilibrium structures form when ${\rm Pe}_{\rm S} > 1$ \cite{fortini:2016, howard:lng:2017}. Last, we have chosen to use $d_{\rm S}$ and $D_{\rm S}$ to define the units of our simulations, and $D_{\rm B}$ is proportional to $D_{\rm S}$ through the particle diameters.

With these constraints and knowledge, we identified three independent parameters to design, $\vv{x} = (d_{\rm B}/d_{\rm S}, \log_{10} {\rm Pe}_{\rm S}, \phi_{0,{\rm S}})$. Based on typical values used in experiments \cite{schulz:2018, schulz:2021}, we bounded these parameters to $1 \leq d_{\rm B}/d_{\rm S} \leq 5$, $0 \leq \log_{10}{\rm Pe}_{\rm S} \leq 2$, and $0.01 \leq \phi_{0, \rm S} \leq 0.19$. We considered $\log_{10}{\rm Pe}_{\rm S}$ instead of ${\rm Pe}_{\rm S}$ because this parameter must span a large range of values. For a given set of parameters, we conducted a simulation using the protocol described in Sec.~\ref{sec:model}. We stopped all simulations when the final total volume fraction $\phi_1 = \phi_0 H_0/H_1$ was $\phi_1 = 0.5$, after which point we do not anticipate significant additional changes in overall film structure. As a result, the final film height was $H_1 = 20\,d_{\rm S}$ for all simulations.

We defined a loss function based on the film composition by volume. We sought to minimize the root mean squared error (RMSE) $\epsilon(\vv{x})$ between the film composition $\nu(z;\vv{x})$ obtained for the set of parameters $\vv{x}$ and a target composition $\nu_0(z)$,
\begin{equation}
    \epsilon^2(\vv{x}) = \frac{1}{H_1}\int_{0}^{H_1}\d{z}\,[\nu(z;\vv{x})-\nu_0(z)]^2.
    \label{eq:rmse}
\end{equation}
In evaluating this integral average, we discarded values of $\nu$ for which the total position-dependent volume fraction was less than $0.005$ because these measurements were less reliable and did not contribute significantly to the overall film structure. $\epsilon$ will be zero if $\nu$ and $\nu_0$ are identical and greater than zero otherwise, so $\epsilon$ should be minimized with respect to $\vv{x}$. Conducting this optimization directly \cite{rios:2013} would be challenging because a new BD simulation needs to be performed to evaluate $\nu$ for each $\vv{x}$. We next discuss the different surrogate models that can be employed to approximate $\epsilon$.

\section{Surrogate models}
\label{sec:surrogate-models}
We considered surrogate models $\hat\epsilon$ for approximating the loss function $\epsilon$ based on both GPR and Chebyshev polynomial interpolation. Both models can be mapped onto the functional form
\begin{equation}
\hat \epsilon(\vv{x}) \approx\sum_{n=1}^{M} c_n \psi_n(\vv{x}),
\label{eq:surrogate}
\end{equation}
where $c_n$ are coefficients and $\psi_n$ are multivariate functions. Such a function can interpolate $M$ values of the loss function $\{\epsilon_1, ..., \epsilon_M\}$ sampled at points $\{\vv{x}_1, ..., \vv{x}_M\}$, and the accuracy of the surrogate model is usually expected to improve when more sample points are used. Each sample point corresponds to one BD simulation. Both types of surrogate models can be evaluated significantly faster than performing a simulation, but there are differences in their functional forms, how data is sampled, and potentially their accuracy for the same number of sample points. We describe the details of the two approaches now, then we will compare their performance in Sec.~\ref{sec:results}.

\subsection{Gaussian process regression}
GPR is a surrogate-modeling strategy that is frequently used in machine learning due to its relative simplicity and versatility \cite{williams:2006, rasmussen:2004, williams:1995, wang_intuitive:2023}. A Gaussian process is defined by a mean function $m(\vv{x})$ for an input $\vv{x}$ and a covariance (or kernel) function $k(\vv{x},\vv{x}')$ between $\vv{x}$ and another input $\vv{x}'$. It is commonly assumed that $m = 0$, e.g., because the mean has been subtracted from the function being approximated. After selecting an appropriate kernel (see below), measurements of the function at a finite number of sample points are used to predict the value of the function at a new point through a probabilistic approach that also estimates the uncertainty in the prediction. GPR can be shown to have the form of eq.~\eqref{eq:surrogate} with $\psi_n(\vv{x}) = k(\vv{x},\vv{x}_n)$ \cite{rasmussen:2004}.

The choice of kernel has a significant impact on the predictive performance of GPR  \cite{genton:2001, duvenaud_automatic_2014}. Kernels are required to have certain properties but commonly used ones are stationary and isotropic, meaning that they depend on only a distance $\xi(|\vv{x}-\vv{x}'|)$ between the inputs \cite{williams:2006}. For this work, we considered multiple kernels: the standard radial basis function (RBF) kernel, the Mat\'{e}rn kernel with exponents 3/2 and 5/2, and the rational quadratic kernel \cite{genton:2001, rasmussen:2004, williams:2006}. Their functional forms are given in Sec.~S1.

For each kernel, we also considered two definitions of the distance: a distance isotropically scaled by a single length $\ell$,
\begin{equation}
\xi(\vv x, \vv x') = \frac{\left| \vv x - \vv x' \right|}{\ell},
\label{eq:scaleddistiso}
\end{equation}
and a distance anisotropically scaled by a vector of lengths $\boldsymbol{\ell}$,
\begin{equation}
\xi(\vv x, \vv x') = \left[\sum_{i=1}^{D}\frac{( x_i - x'_i)^2}{\ell_i^2}\right]^{1/2},
\label{eq:scaleddistaniso}
\end{equation}
where $x_i$ is the $i$-th component of $\vv{x}$ and $D$ is the number of dimensions for the input space ($D=3$ for our problem). The scaling lengths are hyperparameters of the model that can be optimized during model fitting. Anisotropic scaling can produce more accurate predictions than isotropic scaling, but the additional hyperparameters lead to longer training times \cite{soleimani_analyzing_2024, noack_autonomous:2020}.

GPR does not have a required sampling strategy, and initial sample points are often chosen randomly to cover the input space \cite{williams:1995,rasmussen:2004, williams:2006}. These initial samples can be supplemented using adaptive sampling (active learning) strategies, e.g., to reduce the model's uncertainty in specific regions \cite{cohn:1996}. Here, we used Latin hypercube sampling to generate the desired number of sample points. Latin hypercube sampling is a stratified technique that reduces the likelihood of high variability in sample density throughout the domain or an unfavorable bias towards certain dimensions \cite{stein_large_1987}. We used GPy (version 1.13.2) to construct our GPR models \cite{gpy:2014}. The input parameters were linearly transformed to all have the domain $[-1, 1]$, and the values of the loss function being approximated were normalized to have zero mean and unit variance. Complete details on kernel selection and hyperparameter tuning are described in Sec.~S1.  We found that the Mat\'{e}rn kernel with exponent 3/2 and anisotropic distance scaling consistently performed the best out of the kernels we considered (Fig.~S1), so only these results are shown in the main text.

\subsection{Chebyshev polynomial interpolation}
Chebyshev polynomials are a family of orthogonal polynomials that are known to be highly useful for interpolating data \cite{Press:2002}. The interpolant typically uses multivariate basis functions that are products of univariate Chebyshev polynomials of the first kind $\psi_n(\vv{x}) = \prod_{i=1}^D T_{n_i}(x_i)$, where the univariate polynomials associated with each coordinate $T_{n_i}(x_i)$ can have a different degree $n_i$. Unlike GPR, a good choice of sample points is connected to the Chebyshev polynomials that serve as basis functions; here, we used the extrema of the Chebyshev polynomials of the first kind \cite{clenshaw:1960}.

To build our surrogate model, we specified a highest-degree univariate polynomial $T_N$ for all coordinates, so the allowed univariate polynomials for each coordinate were $\{T_0, ..., T_N\}$. The multivariate basis functions were then constructed using a tensor product of univariate basis functions. The highest-degree polynomial has $N+1$ extrema; these were used to construct a corresponding tensor product of multivariate sample points. With this ``dense'' interpolation scheme, there were $M = (N+1)^D$ basis functions and sample points \cite{judd:2014, kieslich:2018}.

Unfortunately, the computational cost of training and evaluating the dense interpolant grows quickly with both $N$ and $D$. Sparse interpolation methods can address this challenge by restricting the sample points and basis functions that are included. Here, we used the Smolyak method \cite{smolyak:1963, judd:2014, kieslich:2018} to select only a subset of the basis functions and sample points from the full tensor product. Briefly, the univariate sample points and basis functions were organized into nested ``levels,'' then the multivariate sample points and basis functions were constructed from tensor products of only certain prescribed levels. More details can be found in Refs.~\citenum{kieslich:2018} and \citenum{judd:2014}. Table \ref{table:smolyak_tensor} compares the total number of sample points and basis functions in both the dense and sparse interpolation schemes for our three-parameter design space. The sparse interpolation scheme reduces the number of points by up to an order of magnitude.

\begin{table}
\caption{Comparison of the number of sample points in dense (tensor product) and sparse (Smolyak sparse product) Chebyshev polynomial interpolation schemes, given a highest-degree $N$ for the univariate polynomials.}
\begin{tabular}{ccc}
$N+1$ & dense & sparse  \\ \hline
3 & 27 & 7 \\
5 & 125 & 25 \\
9 & 729 & 69
\end{tabular}
\label{table:smolyak_tensor}
\end{table}

\section{Results and Discussion}
\label{sec:results}
We assessed the performance of the three different surrogate modeling strategies for our inverse-design problem in two ways. First, we analyzed how well the surrogate model for the loss function $\hat\epsilon$ approximated the true loss function $\epsilon$ (Sec.~\ref{sec:results:error-prediction}). Then, we explored the suitability of the surrogate models for design by minimizing $\hat\epsilon$ with respect to the design parameters and characterizing the film produced by these parameters (Sec.~\ref{sec:results:parameter-prediction}). For both these purposes, we generated 100 points randomly in the design space using Latin hypercube sampling that were not used to train any of the surrogate models and ran a BD simulation for each; these points will be referred to as the target set. We considered dense and sparse versions of three different Chebyshev polynomial interpolants of increasing order (Table~\ref{table:smolyak_tensor}), as well as six GPR models trained using equivalent numbers of sample points taken from Latin hypercube sequences.

\subsection{Approximation of loss function}
\label{sec:results:error-prediction}
To assess the approximation accuracy of the different surrogate models, the film composition for one point in the target set was selected as the design target. The loss between each sample point for the surrogate models and the design target was computed to train the surrogate models. Last, the surrogate models were evaluated at all points in the target set to get the approximated loss $\hat\epsilon$ at each. We also determined the true value of the loss $\epsilon$ at the same points; one point always had $\epsilon=0$ because it was the design target. This procedure was repeated using each point in the target set as the design target to obtain a representative sampling.

\begin{figure}
    \centering
    \includegraphics{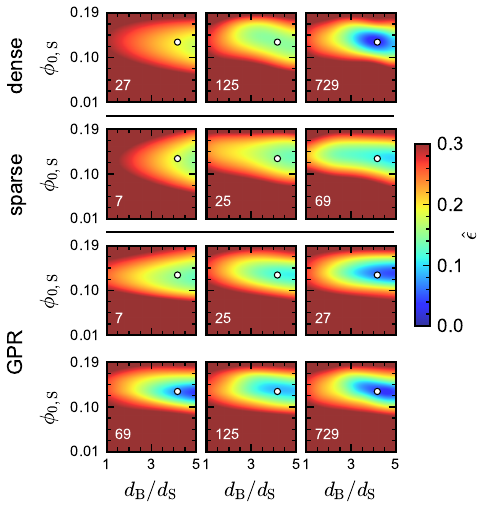}
    \caption{Two-dimensional cross-sections of all surrogate models of loss $\hat{\epsilon}$ for one design target having $d_{\rm B}/d_{\rm S} = 4.1671$, $\log_{10} {\rm Pe}_{\rm S} = 1.5516$ and $\phi_{0, {\rm S}} = 0.1309$. $\log_{10} {\rm Pe}_{\rm S}$ is held constant at the target's value. The rows are grouped into sections by surrogate model type (dense Chebyshev polynomial interpolation, sparse Chebyshev polynomial interpolation, and GPR), and the panels in each section use increasing numbers of sample points (listed in bottom left corner). The design target is marked with a point.}
    \label{fig:heat-maps}
\end{figure}

We first visualized two-dimensional cross-sections of the three-dimensional surrogate models for $\hat{\epsilon}$ for a randomly selected design target (Figs.~\ref{fig:heat-maps}, S2, and S3) to gain a qualitative understanding of the loss function being approximated. We kept one parameter constant at the target's value while varying the other two. Figure \ref{fig:heat-maps} shows the cross-section in the $(d_{\rm B}/d_{\rm S}, \phi_{\rm S})$ plane. We regard the dense Chebyshev polynomial interpolant and the GPR with the largest numbers of points as being highly accurate approximations (see below). The loss function appeared to be convex in these cross-sections, possessing a single minimum. This minimum was in roughly the correct location for both approximations, although the minimum of the dense Chebyshev polynomial interpolant was clearer and closer to the target. The other, less accurate surrogate models qualitatively resembled the more accurate approximations. For example, they all tended to have smaller values in similar ranges of $\phi_{\rm S}$. They all also mostly improved as sample points were added. However, these surrogate models did not have as clear of a minimum and had nonzero values near the design target. This behavior is especially noticeable for the 125-point GPR model, for which the minimum in this cross-section is closer to the right location than for the 69-point GPR model but has a loss that is larger than at the minimum of the 729-point GPR model. For such models, the global minimum of $\hat\epsilon$ may sit at a different value of ${\rm Pe}_{\rm S}$, which may in turn affect the result of using $\hat\epsilon$ for design.

We next constructed parity plots of $\hat\epsilon$ vs.~$\epsilon$ for each three-dimensional surrogate model using the data from all design targets, and we calculated the coefficient of determination $R^2$ (Figs.~S4--S6). As expected, $R^2$ increased for all surrogate-modeling strategies as the number of sample points increased, achieving a best value of 1.00 for both the dense Chebyshev polynomial interpolant and the GPR models and a slightly smaller best value of 0.99 for the sparse Chebyshev polynomial interpolant. We consistently found a somewhat better $R^2$ for the GPR models compared to the Chebyshev polynomial interpolants with equivalent numbers of sample points. The error in most of the surrogate models was more prevalent for smaller values of $\epsilon$, with the surrogate models tending to overpredict $\hat\epsilon$. Because surrogate-based optimization seeks to minimize $\hat\epsilon$, inaccuracy for small values of $\epsilon$ may be important when using the surrogates for design.

\begin{figure}
    \centering
    \includegraphics{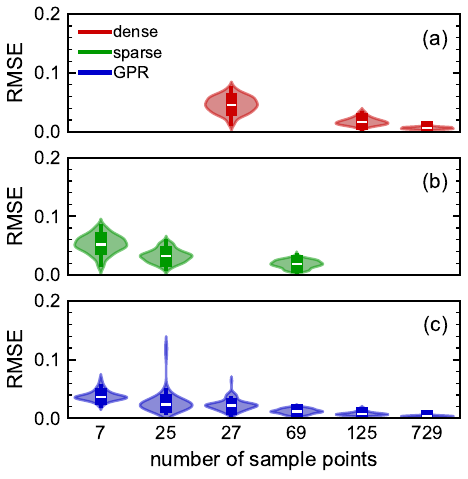}
    \caption{Distribution of RMSE for surrogate models of loss $\hat\epsilon$ for different targets using (a) dense and (b) sparse Chebyshev polynomial interpolation as well as (c) GPR. Quartiles of the distributions are also given in Table S1.}
    \label{fig:violin-plots-rmse}
\end{figure}

We last quantified the performance of the surrogate models across the different design targets by computing the RMSE between the approximated and true values of the loss function at all points in the target set for each. This procedure gave us a distribution of RMSEs for each surrogate model (Fig.~\ref{fig:violin-plots-rmse}). The median RMSE consistently improved for all surrogate models as the number of points increased, and the distributions also tended to narrow. The sparse Chebyshev polynomial interpolants had comparable but somewhat larger median RMSE than their dense counterparts; however, they used substantially fewer sample points. The GPR models consistently had smaller median RMSE than the dense or sparse Chebyshev polynomial interpolants with the same numbers of sample points.  It is notable, though, that the GPR models sometimes had worse maximum errors for small training data sizes (e.g., 25 sample points); this behavior may be due to the random nature of the sampling. 

\subsection{Suitability for design}
\label{sec:results:parameter-prediction}
Having assessed the approximation accuracy of the surrogate models, we proceeded to probe their suitability for inverse design. For each design target, we minimized our surrogate models for the loss $\hat\epsilon$ with respect to the design parameters $\vv{x}$ subject to their box constraints using the L-BFGS-B method. We noted that some of the loss functions (or their approximations) had shallow regions that led to early stopping. To mitigate this issue, we used a heuristic approach to select good initial guesses for the minimization. We first evaluated the surrogate model at $10^5$ random points in the design space using Latin hypercube sampling. We then clustered the $10^3$ points with the smallest values of $\hat\epsilon$ using the \textit{k}-means algorithm to attempt to identify local minima. We used four clusters for this step after confirming we found no significant differences using eight. Last, the closest points to the cluster centroids were used as initial guesses for the L-BFGS-B method. The point having the smallest $\hat\epsilon$ after minimization was taken as the surrogate-based design, $\vv{\hat{x}}^*$.

\begin{figure}
    \centering
    \includegraphics{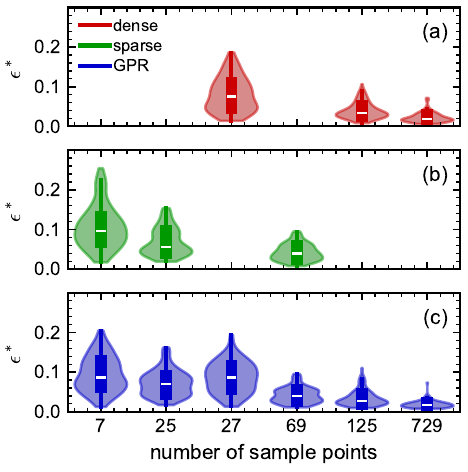}
    \caption{Distribution of true loss $\epsilon^*$ for surrogate-based design for different targets using (a) dense and (b) sparse Chebyshev polynomial interpolation as well as (c) GPR. Quartiles of the distributions are also given in Table S2.}
    \label{fig:opti_gap}
\end{figure}

We then calculated the true loss for the surrogate-based design, $\epsilon^* = \epsilon(\vv{\hat{x}}^*)$, by running an additional BD simulation for each $\vv{\hat{x}}^*$. We expect $\epsilon^*$ to be small if the surrogate-based design is close to the target (where $\epsilon = 0$) and to have larger values for less optimal results. As for the approximation error, we created distributions of $\epsilon^*$ for each surrogate model using all design targets (Fig.~\ref{fig:opti_gap}). The lowest-order dense and sparse Chebyshev polynomial interpolants had broad distributions in $\epsilon^*$ with a median near 0.1, and the equivalent GPR models had comparable behavior. The median value of $\epsilon^*$ systematically decreased for both types of Chebyshev polynomial interpolants as the number of sample points increased. The dense interpolant typically had a better median error than the sparse interpolant for the same order of approximation but required substantially more sample points. The median $\epsilon^*$ was 0.02 and 0.04 for the best dense and sparse Chebyshev polynomial interpolants, respectively. Consistent with Fig.~\ref{fig:violin-plots-rmse}, the GPR models with equivalent numbers of points to these models had a somewhat smaller median value and narrower distribution of $\epsilon^*$, indicating a somewhat better overall quality of design.

\begin{figure}
    \centering
    \includegraphics{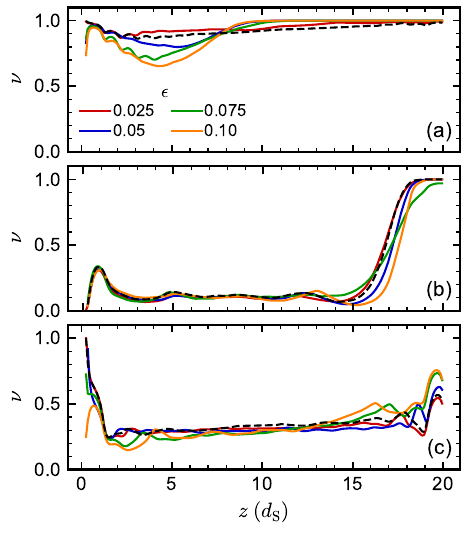}
    \caption{Composition by volume $\nu$ with varying loss $\epsilon$ relative to a target $\nu_0$, selected from the sample points of the largest dense Chebyshev polynomial interpolant. The three targets have $(d_{\rm B}/d_{\rm S}, \log_{10} {\rm Pe}_{\rm S}, \phi_{0,{\rm S}})$ of (a) $(2.3762, 0.3932, 0.1857)$, (b) $(4.6304, 1.6225, 0.0404)$, and (c) $(2.0931, 0.2257, 0.0691)$.}
    \label{fig:error_weighted_vol_frac}
\end{figure}

To contextualize the performance of the various surrogate models, we compared compositions by volume $\nu$ with varying losses $\epsilon$ to the design target $\nu_0$ (Fig.~\ref{fig:error_weighted_vol_frac}). We selected profiles having errors closest to the desired values from the sample points of the largest dense Chebyshev polynomial interpolant for three different targets. For all three targets, good agreement was found between $\nu$ and $\nu_0$ when $\epsilon = 0.025$. When $\epsilon$ increased to $0.05$, there was less consistency between targets: significant differences between $\nu$ and $\nu_0$ were apparent for the first target [Fig.~\ref{fig:error_weighted_vol_frac}(a)] but not for the other two [Fig.~\ref{fig:error_weighted_vol_frac}(b)--(c)]. There were apparent differences between $\nu$ and $\nu_0$ for all targets at larger $\epsilon$.

We hence consider $\epsilon^* < 0.025$ to be good, but $\epsilon^* < 0.05$ may be acceptable for some targets. All surrogate models were able to achieve median $\epsilon^* < 0.05$, requiring 125 sample points for dense Chebyshev polynomial interpolation and 69 sample points for sparse Chebyshev polynomial interpolation and GPR. The dense Chebyshev polynomial interpolation and GPR models also both achieved median $\epsilon^* < 0.025$ but required 729 sample points to do so. The sparse Chebyshev polynomial interpolant did not achieve median $\epsilon^* < 0.025$ but might do so by increasing the order of approximation further. The next sparse Chebyshev polynomial interpolant, having $N+1 = 17$, would require 177 sample points, which is still significantly less than used by the largest dense interpolant considered here.

\section{Conclusions}
\label{sec:conclusions}
We have developed and tested a surrogate-modeling framework for designing the drying-induced assembly of multicomponent colloidal-particle films. Specifically, we considered the assembly of bidisperse hard-sphere colloidal particles. The design problem was formulated by minimizing the loss in the targeted composition by volume in the dried film with respect to three physically motivated and constrained parameters: the particle size ratio, the P\'{e}clet number for the smaller particles (drying rate), and the initial volume fraction of small particles. We critically assessed the use of three different types of surrogate models, one based on GPR and the others based on dense and sparse Chebyshev polynomial interpolation, for approximating and minimizing the loss. Overall, we found surrogate modeling to be highly effective for addressing this inverse design problem, and we obtained good results using surprisingly few sample points to train them. GPR typically gave somewhat better approximations and designs than either Chebyshev polynomial interpolant. The sparse Chebyshev polynomial interpolants performed competitively with their dense equivalents, particularly when considering the significant difference in number of sample points. Practically, we noted that the Chebyshev polynomial interpolants were significantly easier and faster to train than the GPR models because they did not require numerical hyperparameter optimization.

In real colloidal films, there are particle properties that may need to be accounted for that we have not considered here, including Hamaker constants (van der Waals forces), surface charge (electrostatics), particle shape, and presence of surfactants or counterions in solution, as well as processing properties such as film height and initial particle concentration that are not necessarily constrained. The inclusion of these parameters in the design space increases its dimensionality significantly, necessitating efficient use of data. In this context, we regard the sparse Chebyshev polynomial interpolants to strike a promising balance between accuracy and practical convenience.

\section*{Supplementary Material}
See the supplementary material for details on kernel selection and hyperparameter tuning for GPR models, additional cross-sections of surrogate models for one design target, parity plots of loss function for all surrogate models, and statistics for the distributions of RMSE and $\epsilon^*$ for all surrogate models.

\section*{Conflicts of interest}
The authors have no conflicts to disclose.

\section*{Data Availability}
The data that support the findings of this study are available from the authors upon reasonable request.

\section*{Acknowledgments}
This material is based upon work supported by the National Science Foundation under Award No.~2442526 and by the Auburn University Research Support Program. This work was completed with resources provided by the Auburn University Easley Cluster.

\bibliography{references}
\end{document}


\title{Supplementary material for ``Inverse design of drying-induced assembly of multicomponent colloidal-particle films using surrogate models''}

\author{Mayukh Kundu}
\affiliation{Department of Chemical Engineering, Auburn University, Auburn, AL 36849, USA}

\author{Michaela Bush}
\affiliation{Department of Chemical Engineering, Auburn University, Auburn, AL 36849, USA}

\author{Chris A. Kieslich}
\affiliation{Wallace H. Coulter Department of Biomedical Engineering, Georgia Institute of Technology, Atlanta, Georgia 30332, United States}

\author{Michael P. Howard}
\email{mphoward@auburn.edu}
\affiliation{Department of Chemical Engineering, Auburn University, Auburn, AL 36849, USA}

\maketitle

\section{GPR kernel selection and hyperparameter optimization}
Four kernels were considered for GPR: the radial basis function (RBF), the Mat\'{e}rn (M) kernel, and the rational quadratic (RQ) kernel. The RBF kernel is
\begin{equation}
\label{eq:RBFkernel}
k_{\rm RBF}(\xi) = \sigma^{2}_{k}\exp\left(-\frac{\xi^{2}}{2}\right),
\end{equation}
where $\sigma^{2}_{k}$ is the kernel variance hyperparameter and $\xi$ is the scaled distance as defined in the main text. The Mat\'ern kernel with exponent $\nu$ is
\begin{equation}
\label{eq:Maternkernel}
k_{\rm M}(\xi) = \sigma^{2}_{k}\frac{2^{1-\nu}}{\Gamma(\nu)}\left( \sqrt{2\nu}\xi \right)^{\nu}K_{\nu}\left( \sqrt{2\nu}\xi \right),
\end{equation}
where $K_{\nu}$ is the modified Bessel function of the second kind. We considered the commonly used exponents $3/2$ and $5/2$, which we denote as M32 and M52, respectively. For these exponents, $k_{\rm M}$ simplifies to
\begin{align}
k_{{\rm M}32}(\xi) &= \sigma^{2}_{k}\left( 1 + \sqrt{3}\xi \right)\exp \left(-\sqrt{3}\xi  \right) \label{eq:Matern32kernel}, \\
k_{{\rm M}52}(\xi) &= \sigma^{2}_{k}\left( 1 + \sqrt{5}\xi + \frac{5\xi^{2}}{3}\right)\exp \left(-\sqrt{5}\xi  \right)
\label{eq:Matern52kernel}
.
\end{align}
Last, the RQ kernel is
\begin{equation}
\label{eq:RQkernel}
k_{\rm RQ}(\xi) = \sigma^{2}_{k}\left( 1 + \frac{\xi^{2}}{2\alpha} \right)^{-\alpha},
\end{equation}
where $\alpha$ is the scale mixture hyperparameter.

\begin{figure}[!h]
  \centering
  \includegraphics{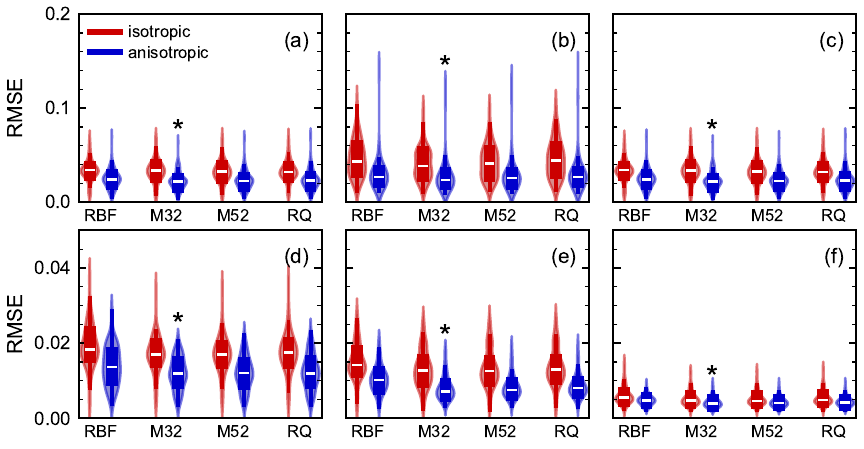}
  \caption{Distributions of RMSE for GPR models of loss $\hat\epsilon$ for different targets using different kernels with isotropic (red) or anisotropic (blue) length scaling trained on varying numbers of sample points: (a) 7, (b) 25, (c) 27, (d) 69, (e) 125 and (f) 729.}
  \label{fig:gpmviolinrmse}
\end{figure}

Once a kernel is selected, its hyperparameters can be numerically optimized using a Bayesian approach.\textsuperscript{52} We performed a preliminary analysis on a limited number of GPR models to explore the effects of using different initial guesses for the kernel hyperparameters and the number of times the hyperparameter optimization should be restarted. We found that good, consistent results were obtained with reasonable computational cost using initial guesses of $\ell_i = 100$ for the length scale in each dimension $i$, $\sigma_k^2 = 1$, and $\alpha = 1$  (all quantities implicitly in correct units, as needed) and ten restarts of the optimization. We then analyzed the distribution of RMSE for approximating the loss using all four kernels with isotropic or anisotropic length scaling and different numbers of sample points (Fig.~\ref{fig:gpmviolinrmse}). We found that anisotropic length scaling consistently gave a smaller median RMSE than isotropic length scaling, and the M32 kernel with anisotropic length scaling consistently had the smallest median RMSE. We hence used this kernel for all the results and analysis presented in the main text.

\FloatBarrier
\section{Supplementary figures and data}
\begin{figure}[!h]
  \centering
  \includegraphics{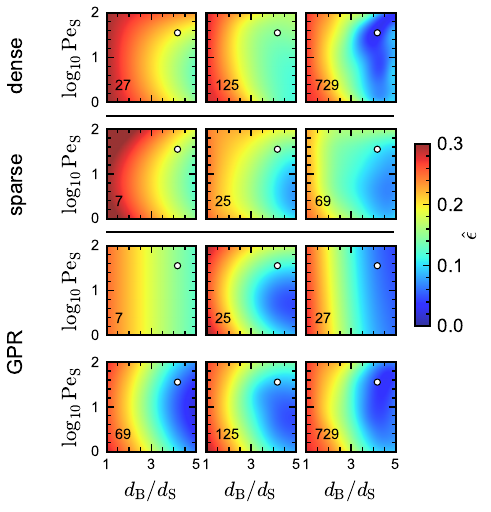}
  \caption{Same as Fig.~4 but with $\phi_{0,{\rm S}}$ held at the target's value.}
\end{figure}

\begin{figure}[!h]
  \centering
  \includegraphics{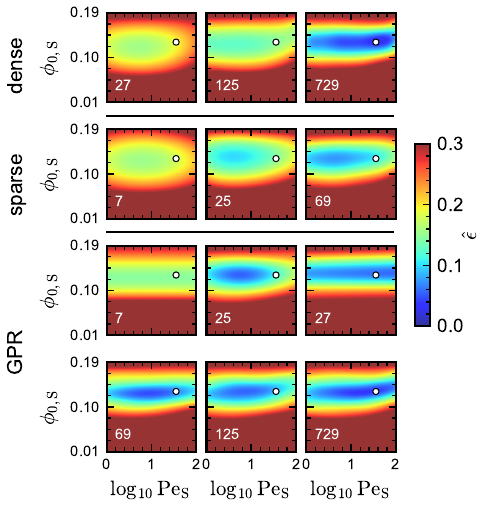}
  \caption{Same as Fig.~4 but with $d_{\rm B}/d_{\rm S}$ held at the target's value.}
\end{figure}

\begin{figure}[!h]
  \centering
  \includegraphics{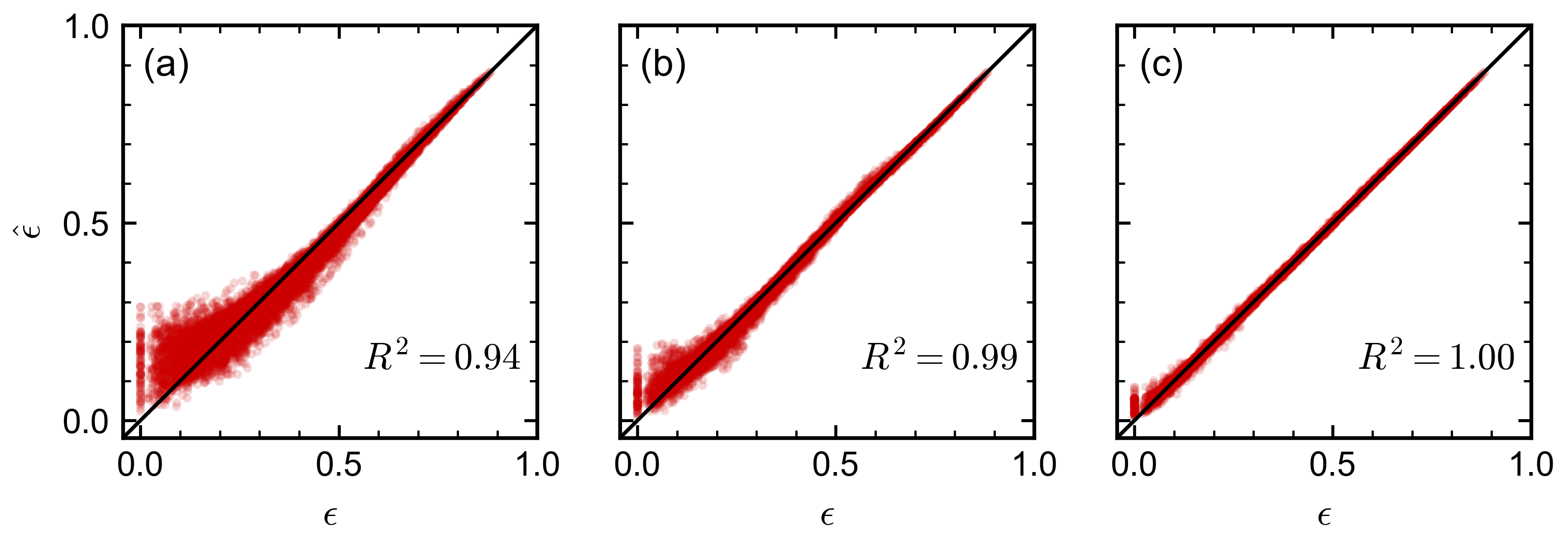}
  \caption{Approximate loss $\hat{\epsilon}$ vs.~true loss $\epsilon$ for dense Chebyshev polynomial interpolants trained using (a) 27, (b) 125, and (c) 729 sample points. The coefficient of determination $R^2$ is given for each surrogate model.}
  \label{fig:parity:dense}
\end{figure}

\begin{figure}[!h]
  \centering
  \includegraphics{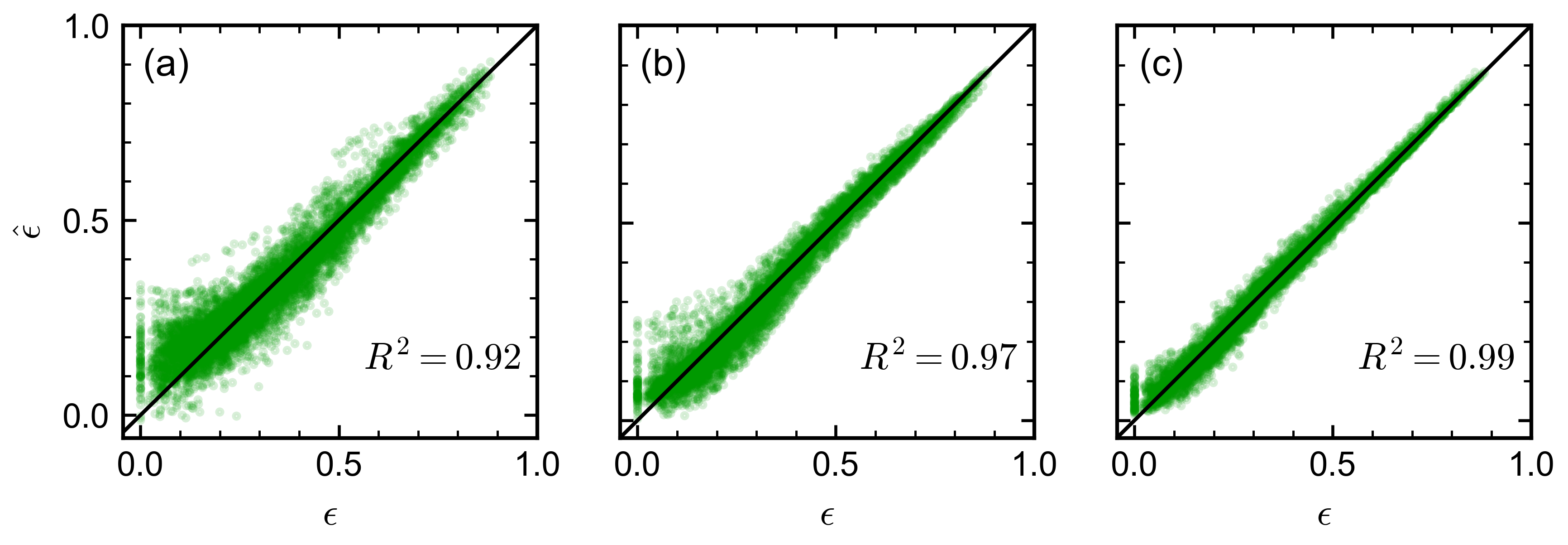}
  \caption{Same as Fig.~\ref{fig:parity:dense} but for sparse Chebyshev polynomial interpolants trained using (a) 7, (b) 25, and (c) 69 sample points.}
\end{figure}

\begin{figure}[!h]
  \centering
  \includegraphics{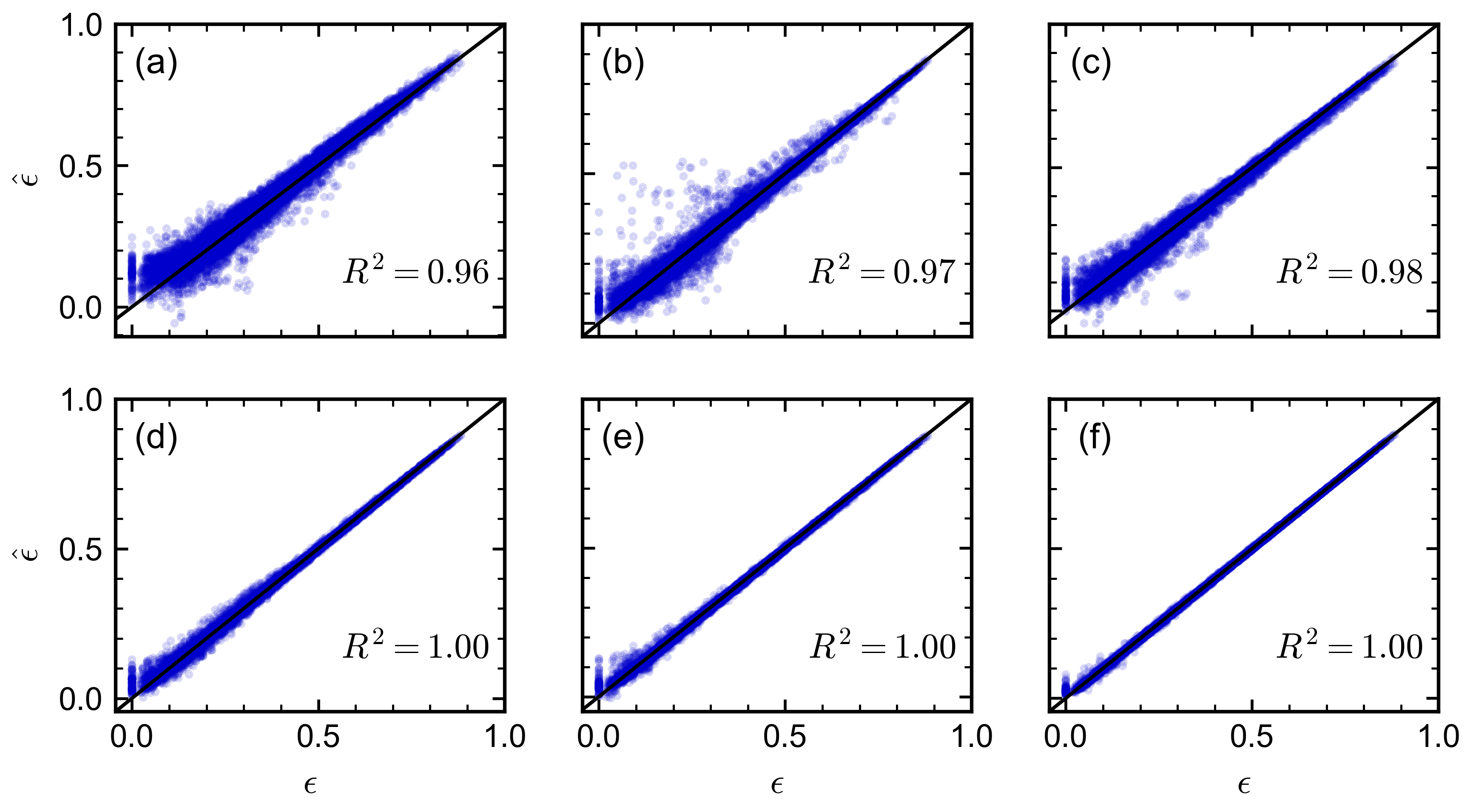}
  \caption{Same as Fig.~\ref{fig:parity:dense} but for GPR models trained using (a) 7, (b) 25, (c) 27, (d) 69, (e) 125, and (f) 729 sample points.}
\end{figure}

\begin{table}[!h]
\caption{Lower quartile $Q_1$, median $Q_2$, and upper quartile $Q_3$ of the distributions shown in Fig.~3.}
\begin{tabularx}{5in}{CCCCC}
& number of sample points & $Q_1$ & $Q_2$ & $Q_3$ \\ \hline
\multirow{3}{*}{sparse} & 7                       & 0.044          & 0.052  & 0.063          \\ 
                        & 25                      & 0.024          & 0.032  & 0.039          \\ 
                        & 69                      & 0.014          & 0.019  & 0.025          \\ \hline
\multirow{3}{*}{dense}  & 27                      & 0.036          & 0.457  & 0.057          \\ 
                        & 125                     & 0.013          & 0.016  & 0.057          \\ 
                        & 729                     & 0.005          & 0.006  & 0.008          \\ \hline
\multirow{6}{*}{GPR}    & 7                       & 0.033          & 0.036  & 0.042          \\
                        & 25                      & 0.019          & 0.024  & 0.031          \\ 
                        & 27                      & 0.016          & 0.022  & 0.025          \\ 
                        & 69                      & 0.009          & 0.019  & 0.015          \\  
                        & 125                     & 0.006          & 0.007  & 0.009          \\ 
                        & 729                     & 0.003          & 0.004  & 0.005 \\ \hline
\end{tabularx}
\label{tab:rmse-distrib}
\end{table}

\begin{table}[!h]
\caption{Same as Table~\ref{tab:rmse-distrib} but for the distributions shown in Fig.~4.}
\begin{tabularx}{5in}{CCCCC}
 & number of sample points & $Q_1$ & $Q_2$ & $Q_3$ \\ \hline
\multirow{3}{*}{sparse} & 7                       & 0.068          & 0.096  & 0.132          \\ 
                        & 25                      & 0.040          & 0.056  & 0.095          \\ 
                        & 69                      & 0.024          & 0.039  & 0.057          \\ \hline
\multirow{3}{*}{dense}  & 27                      & 0.045          & 0.076  & 0.111          \\ 
                        & 125                     & 0.026          & 0.034  & 0.051          \\ 
                        & 729                     & 0.014          & 0.018  & 0.029          \\ \hline
\multirow{6}{*}{GPR}    & 7                       & 0.063          & 0.087  & 0.127          \\ 
                        & 25                      & 0.044          & 0.071  & 0.091          \\ 
                        & 27                      & 0.059          & 0.086  & 0.116          \\ 
                        & 69                      & 0.027          & 0.040  & 0.055          \\ 
                        & 125                     & 0.020          & 0.028  & 0.046          \\ 
                        & 729                     & 0.012          & 0.017  & 0.023          \\ \hline
\end{tabularx}
\end{table}